\documentclass{ws-procs9x6}

\begin{document}

\title{$INTEGRAL$ three years later}

\author{L.~FOSCHINI, G.~DI~COCCO, G.~MALAGUTI}

\address{INAF/IASF, Sezione di Bologna\\
Via Gobetti 101, 40129 - Bologna (Italy)\\ 
E-mail: foschini, dicocco, malaguti @bo.iasf.cnr.it}

\maketitle

\abstracts{The ESA \emph{INTErnational Gamma-Ray Astrophyisics Laboratory} (\emph{INTEGRAL}) is composed 
of two main instruments (IBIS and SPI) for the $\gamma-$ray astrophysics ($20$~keV -- $10$~MeV), plus two 
monitors (JEM-X and OMC) for X-ray and optical counterparts. It was launched on October $17^{\rm th}$, $2002$, 
from the Baikonur cosmodrome (Kazakhstan) and it will be active at least up to $2008$. 
A selection of its main scientific contributions obtained to date are presented.}

\section{Introduction}
The \emph{INTE}rnational \emph{G}amma-\emph{R}ay \emph{A}strophysics \emph{L}aboratory (\emph{INTEGRAL})\cite{WINKLER} 
is an observatory of the \emph{European Space Agency} (ESA). Launched in 
$2002$, with a nominal duration of $2$ years, it has already been extended up to $2008$. \emph{INTEGRAL} has 
been built by the ESA member states with the participation of United States, Russia, Czech Republic, and Poland. 
\emph{Alenia Spazio} (Italy) has been appointed by ESA as prime contractor for the design, integration, and testing 
of the satellite\cite{JENSEN}. The \emph{Mission Operation Centre} (MOC) at Darmstadt (Germany) is continuously 
in touch with the satellite through two ground stations, one of ESA located at Redu (Belgium) and the other of 
NASA at Goldstone (USA). The \emph{INTEGRAL Science Operation Centre} (ISOC) was initially at ESA ESTEC 
(The Netherlands), and in $2005$ moved to ESAC in Spain\cite{MUCH}. The \emph{INTEGRAL Science Data Centre} (ISDC) is located at Versoix in Switzerland\cite{ISDC}. 

Two are the main instruments on board: SPI (\emph{SPectrometer on Integral})\cite{SPI} and IBIS (\emph{Imager on Board Integral Satellite})\cite{IBIS}. The first is dedicated to high-energy resolution spectroscopy ($2.5$~keV at $1.3$~MeV) and
it is composed of $19$ Germanium detectors operating in the $20$~keV -- $8$~MeV energy range. The second is an imaging telescope optimized for high-resolution ($12'$) wide field of view (FOV, $29^{\circ}\times 29^{\circ}$) point source reconstruction with moderate energy resolution ($8$\% at $100$~keV and $10$\% at $1$~MeV). IBIS is composed of two layers: ISGRI (\emph{INTEGRAL Soft Gamma-Ray Imager}\cite{ISGRI}), a $128\times 128$ pixels CdTe detector organized in $8$~modules operating in the $20$~keV -- $1$~MeV energy range, and PICsIT (\emph{Pixellated Imaging CaeSium Iodide Telescope}\cite{PICSIT}), a $64\times 64$ pixels CsI detector organized in $8$~modules operating in the $175$~keV -- $10$~MeV energy range. 

These instruments are complemented by two monitors: JEM-X (\emph{Joint European Monitor X-rays}\cite{JEMX}), operating in the $3-35$~keV energy band with angular resolution of $3'$, and OMC (\emph{Optical Monitor Camera}\cite{OMC}), for optical study in the Johnson V band with magnitude limit $18$. 

IBIS, SPI, and JEM-X detectors are all coupled with tungsten coded-masks: SPI and JEM-X make use of a hexagonal uniformly redundant array (HURA), while the IBIS mask is a modified uniformly redundant array (MURA). Specific mathematical operations (deconvolution) are needed to obtain images, spectra, and lightcurves, and are described in Goldwurm et al.\cite{IBIS1} and Gros et al.\cite{IBIS2} for IBIS, in Skinner \& Connell\cite{SPI1} and Strong\cite{SPI2} for SPI, and in Westergaard et al.\cite{JEMXSW} for JEM-X.

Since \emph{INTEGRAL} is an observatory, most of its time is dedicated to the astronomical community, through 
annual announcements of opportunity (AO). However, part of the observing time is reserved to the Instrument Teams 
and constitutes the Core Programme (CP). The CP covered $35$\% of the total observing time during the first year 
and decreased to $30$\% in the second year, and to $25$\% in the third year. The scientific topics of the CP are 
focused on regular scans of the Galactic plane and deep exposures on the central radian of the Milky Way and other 
specific regions (Norma and Scutum Arms, Vela and Virgo regions). One year after the observation, all the data 
become public\footnote{For more details, see the documentation enclosed to each Announcement of Opportunity available at \texttt{http://integral.esac.esa.int/}.}. 

\section{Science}
The main scientific topics addressed by \emph{INTEGRAL} satellite observations can be divided into: \emph{Compact objects:}  neutron stars, anomalous X-ray pulsars, soft-gamma repeaters, black holes; \emph{The Milky Way:} the Galactic Centre and Sgr A*, diffuse and line emission; \emph{Stellar nucleosynthesis:} hydrostatic (AGB, Wolf-Rayet) and explosive (SN, Novae); \emph{Extragalactic astrophysics:} active galactic nuclei (seyferts, blazars, radiogalaxies), clusters of galaxies, gamma-ray bursts (GRB); \emph{Identification of high-energy sources:} EGRET unidentified sources, new discoveries; \emph{Unexpected discoveries}. 

In the following we shortly present a selection of the most important results obtained with \emph{INTEGRAL}, with apologies to those about which we cannot report, because of space limitations.

\subsection{The Galactic Centre}
One of the most important results was obtained thanks to the unprecedented angular resolution of IBIS/ISGRI. The problem 
was the understanding of the excess of diffuse emission in the Galactic centre: indeed, previous missions found a 
strong $\gamma-$ray diffuse emission toward the centre of the Milky Way, that was thought to be originated from the 
interaction of cosmic-rays with interstellar medium (see Pohl\cite{POHL} for a review). However, this model cannot 
explain excesses in the spectrum around $100$~keV and above $1$~GeV. \emph{INTEGRAL} performed a deep exposure ($1.5$~Ms) 
toward the central radian of the Galaxy, and demonstrated that the excess was due to point sources not resolved by 
previous experiments\cite{LEBRUN}.

Again IBIS/ISGRI performed an accurate study of Sgr~A* and found a new hard X-ray source IGR~J$1745.6-2901$, consistent 
with the Galactic Centre within $\approx 1'$, so that the association with Sgr~A* is not conclusive\cite{SGRA}. On the 
other hand, the flux of IGR~J$1745.6-2901$ is not consistent with the diffuse and point sources emission within $10'$ 
extrapolated from \emph{Chandra} and \emph{XMM-Newton} observations. These two X-ray satellites have observed some 
transients in the ISGRI error box, but these are too weak and soft to be associated with the hard X-ray source. 

Another newly discovered \emph{INTEGRAL} source, namely IGR~J$17475-2822$, has been associated with the giant molecular cloud Sgr~B2\cite{REV}. To explain how a molecular cloud could emit hard X-rays, it has been suggested that the radiation could be the result of a past activity of Sgr~A* ($\approx 300-400$ years ago, when it behaved as an active galactic nucleus), that was Compton scattered and reprocessed by Sgr~B2\cite{REV}.

The spectrometer SPI focused on the electron-positron annihilation line ($511$ keV) and found a significant ($\approx 50\sigma$) emission from the Galactic bulge\cite{511}. This emission region is centered on the Galactic centre, is symmetric, and is about $8^{\circ}$ wide. There is no evidence of the positive latitude enhancement found by OSSE\cite{OSSE} and the emission along the Galactic disk is constrained with low significance ($\approx 4\sigma$).

\subsection{Compact objects}
Thanks to its characteristics, \emph{INTEGRAL} can give important contribution mainly in the astrophysics of compact 
objects. Indeed, with a sensitivity of about $(0.8-1)\times 10^{-10}$~erg~cm$^{-2}$~s$^{-1}$ in the $20-40$~keV energy 
band ($10-15$~mCrab) for a $3\sigma$ detection in $2$~ks, the IBIS/ISGRI detector is efficiently used to monitor the 
X-ray binaries population in the Milky Way with the regular scans of the Core Programme. On December $2$, $2004$, 
during these scans, it was discovered the fastest millisecond pulsar ever known: it is the new source IGR~J$00291+5934$, 
that has a pulsation period of $1.67$~ms\cite{SHAW}. In December $2004$, \emph{INTEGRAL} was the first, of more than 
twenty satellites, to detect a gigantic flare from the soft $\gamma-$ray repeater SGR~$1806-20$, so intense to 
significantly ionize the Earth's upper atmosphere\cite{SGR1806,SGRACS,HURLEY}. 

The high angular resolution of \emph{INTEGRAL} in the hard X-rays results crucial to disentangle the emission of 
different sources in crowded fields: for example, Paizis et al.\cite{ADA} succeeded in studying GX~$5-1$, that is 
apparently located close ($40'$) to the black hole GRS~$1758-258$. Another case is represented 
by 4U~$1630-47$, a black hole in the Norma Arm: Tomsick et al.\cite{4U} have shown that the extreme behaviour during 
the period $2002-2004$ is not in agreement with the classical definitions of spectral states.

Finally, among black holes, Cyg~X-1 has been extensively studied during the Performance Verification phase, as 
it was the selected target for the first light of the main instruments\cite{CX1SPI,CX1INTRXTE,CX1IBIS}.

\subsection{New sources and survey}
On January $29$, $2003$ \emph{INTEGRAL} discovered its first new source\cite{IGR}, that was called IGR~J$16318-4848$. 
It is located in the Norma Arm of the Galaxy, an active star forming region, where several other sources have been found 
later\footnote{See the web page on the new \emph{INTEGRAL} sources maintained by J. Rodriguez at 
\texttt{http://isdc.unige.ch/$\sim$rodrigue/html/igrsources.html}}. The high energy spectrum, from soft to hard X-rays 
as result of a joint observation of \emph{XMM-Newton} and \emph{INTEGRAL}, together with optical data, suggest that 
IGR~J$16318-4848$ could be an accreting neutron star with a high-mass companion, enshrouded by a Compton-thick 
environment ($N_{\rm H}\approx 10^{24}$~cm$^{-2}$), perhaps the progenitor of a new population of 
X-ray binaries\cite{RW}. See also Kuulkers\cite{KULK} for a recent review on this type of sources.

Recently, one of the new \emph{INTEGRAL} sources, IGR~J$18135-1751$, has been associated with an unidentified TeV
source discovered by the HESS telescope\cite{HESS}, namely HESS~J$1813-178$. The spectral energy distribution suggests
that the object could be a pulsar wind nebula embedded in its supernova remnant\cite{UBERT}. Soon after, \emph{INTEGRAL}
discovered the soft $\gamma-$ray emission from another unidentified TeV source (HESS~J$1837-069$), that appears to be
another pulsar or supernova remnant\cite{MAL}. 

Several other new sources have been discovered by \emph{INTEGRAL} up to date: the first IBIS/ISGRI source 
catalog\cite{BIRD}, referring to the data from February to October $2003$, contains $123$ known and $28$ new sources 
down to a flux limit of $\approx 1.7\times 10^{-11}$~erg~cm$^{-2}$~s$^{-1}$ in the $20-100$~keV energy range ($1$~mCrab). 
The latest publicly available update back to a presentation by A. Bird\footnote{Talk given at the 
\emph{Internal INTEGRAL Science Workshop} held at ESA/ESTEC on $18-21$ January 2005. 
See \texttt{http://integral.esac.esa.int/workshops/Jan2005/}.} in January 2005 and reported $229$ known and $31$ new sources.

\subsection{Gamma-Ray Bursts (GRB)}
\emph{INTEGRAL} has an elongated orbit with $72$~hours period. This, in addition to the availability of the two ground 
stations of Redu and Goldstone, allows a continuous monitoring of the data in near-real time. This is performed by the 
\emph{INTEGRAL Burst Alert System} (IBAS\cite{IBAS}), in order to distribute the GRB coordinates to the astronomical 
community with unprecedented coupling of high accuracy ($\approx 3'$) and low time delay ($\approx 20-30$~s). See 
Mereghetti \& G\"otz\cite{IBAS2} for a review of the IBAS activity. 
Among the several GRB detected by \emph{INTEGRAL}, it is worth mentioning the first that was in the field of view 
of the instruments (IBIS, SPI). It occurred on November $25$, $2002$, during the Performance Verification phase\cite{BAZPAZ,MALAGUTI}. Moreover, \emph{INTEGRAL} detected also the GRB nearest to the Earth ($z=0.106$)\cite{SAZONOV}.

Other efficient ways to detect GRB (particularly out of the instruments FOV) are available on board \emph{INTEGRAL} satellite with the Anticoincidence Shield (ACS) of SPI\cite{RAU}, the IBIS Compton mode (that makes use of both layers ISGRI and PICsIT)\cite{MARC1}, and the spectral timing mode of IBIS/PICsIT\cite{MARC2}. In these ways, it is possible to reach energies up to a few MeV. 

\subsection{Active Galactic Nuclei}
Although AGN are numerically much less than other sources detected by \emph{INTEGRAL}, there are some interesting 
studies worth mentioning. Six blazars have been detected by \emph{INTEGRAL} to date and it is worth noting that these 
are all also EGRET sources (i.e., that have been detected at energies greater than $100$~MeV by the EGRET experiment 
on board the \emph{Compton Gamma-Ray Observatory}). Three of these six AGN have been detected during a multiwavelength 
Target-of-Opportunity (ToO) programme, led by E. Pian, organized to study the blazars in outburst. 
These are: S5~$0716+714$\cite{PIAN}, following an outburst started at the end of March~$2004$; 3C~$454.3$\cite{3C454}, recently observed during the long outburst of May~$2005$; and S5~$0836+710$, serendipitously detected in the field of view of the observation of S5~$0716+714$\cite{PIAN}. The remaining three blazars are the well known 3C~$273$\cite{3C273} and 3C~$279$\cite{3C279} in the Virgo region, and PKS~$1830-211$\cite{PKS}, the farthest object ($z=2.51$) detected to date by \emph{INTEGRAL}.

Still in the field of view of the ToO on S5~$0716+714$ are present two more AGN: the Seyfert active nuclei Markarian~3 
and Markarian~6\cite{PIAN}. The data from Mkn~3 show a hint for the presence of a break in the spectrum between $60$ and 
$100$~keV, but the low statistics did not allow to better constrain the value\cite{PIAN}.

Among the Seyferts, other studies have been done on NGC~4388\cite{4388}, joint with \emph{XMM-Newton}, and 
GRS~$1734-292$\cite{GRS1}, a luminous Seyfert 1 behind the Galactic Centre. This latter AGN is also the only source 
detected by \emph{INTEGRAL} inside the $99$\% probability contours of the EGRET unidentified source 3EG~J$1736-2908$ 
and thus opening the possibility that the AGN could be the counterpart of the EGRET source\cite{GRS2}. However, 
since no other Seyferts are known to emit at $E>100$~MeV, if this association will be confirmed by 
future observations (with \emph{GLAST}, for example) it remains to understand what is the physical 
mechanism able to generate high-energy $\gamma-$rays from a Seyfert active nucleus.

Other EGRET AGN observed by \emph{INTEGRAL} are the radio galaxies NGC~6251\cite{NGC6251} and Centaurus~A\cite{ROT}. 
In the former case, a faint doubtful detection was strengthened by the agreement of the measured flux with the spectral energy distribution expected from a synchrotron self-Compton model.

Several other AGN, together with a detection in the $30-50$~keV energy band of the Coma Cluster, are present in the field of view analyzed by Krivonos et al.\cite{KRIVONOS}. This allowed to evaluate the extragalactic hard X-ray background as observed by \emph{INTEGRAL}, but given the small sky area covered ($3$\%), much larger samples are needed to better asses the source contribution in this energy band.

\section{Final Remarks}
Thanks to its main characteristics in the hard X/soft-$\gamma$ rays (high sensitivity, high angular resolution, 
high spectral resolution, wide FOV), at the end of the third year of operation \emph{INTEGRAL} has significantly 
contributed to the advancement of the $\gamma-$ray astrophysics with several important contributions, a few of 
these have been shortly reviewed here.  In addition, a public data archive is available to the astronomical 
community\footnote{Everything for the data analysis (data, software, documentation) can be found at the 
\emph{INTEGRAL Science Data Centre} web page \texttt{http://isdc.unige.ch/}.}, where gigabytes of data are 
still waiting to be analyzed. We hope that this short review can be useful to show the huge work done to 
date, the work still to be done, and the potentialities of \emph{INTEGRAL} in the exploration of the $\gamma-$ray sky.

\end{document}